\documentclass[a4paper,twocolumn,prl]{revtex4}%
\usepackage[colorlinks,linkcolor=blue,urlcolor=blue,citecolor=blue]{hyperref}

\usepackage[latin9]{inputenc}
\usepackage{amsmath}
\usepackage{amssymb}
\usepackage{graphicx}
\usepackage{comment}
\usepackage{mathrsfs}
\usepackage{amsfonts}
\usepackage{float}%
\usepackage{multirow}
\providecommand{\U}[1]{\protect\rule{.1in}{.1in}}
\makeatletter

\DeclareFontEncoding{LGR}{}{}
\DeclareTextSymbol{\~}{LGR}{126}
\@ifundefined{textcolor}{}
{
\definecolor{BLACK}{gray}{0}
\definecolor{WHITE}{gray}{1}
\definecolor{RED}{rgb}{1,0,0}
\definecolor{GREEN}{rgb}{0,1,0}
\definecolor{BLUE}{rgb}{0,0,1}
\definecolor{CYAN}{cmyk}{1,0,0,0}
\definecolor{MAGENTA}{cmyk}{0,1,0,0}
\definecolor{YELLOW}{cmyk}{0,0,1,0}
}

\makeatother
\begin{document}
\title{Flat Bands in Twisted Bilayers of Two-Dimensional Polar Materials}
\author{Xing-Ju Zhao$^{1,2}$}
\thanks{X.J.Z. and Y.Y. contributed equally to this work.}
\author{Yang Yang$^{1}$}
\thanks{X.J.Z. and Y.Y. contributed equally to this work.}
\author{Dong-Bo Zhang$^{2,1}$}
\thanks{Corresponding author}
\email[]{dbzhang@bnu.edu.cn}
\author{Su-Huai Wei$^{1}$}
\thanks{Corresponding author}
\email[]{suhuaiwei@csrc.ac.cn}
\affiliation{$^{1}$Beijing Computational Science Research Center, Beijing 100193, P.R. China}
\affiliation{$^{2}$College of Nuclear Science and Technology, Beijing Normal University, Beijing 100875, P.R. China}

\begin{abstract}
The existence of Bloch flat bands provides an facile pathway to realize strongly correlated phenomena in materials. Using density-functional theory and tight-binding approach, we show that the flat bands can form in twisted bilayer of hexagonal boron nitride ($h$BN). However, unlike the twisted graphene bilayer where a magic angle is needed to form the flat band, for the polar $h$BN, the flat bands can appear as long as the twisted angle is less than certain critical values. Our simulations reveal that the valence band maximum (conduction band minimum) states are predominantly resided in the regions of the moir\'{e} supperlattice where the anion N (cation B) atoms in both layers are on top of each other.  The preferential localization of these valence and conduction states originate from the chemical potential difference between N and B and is enhanced by the stacking effects of N and B in both layers, respectively, as demonstrated by an analysis of the energy level order of the $h$BN bilayers with different stacking patterns.  When these states are spatially localized because regions with a specific stacking pattern are isolated for moir\'{e} supperlattices at sufficient small twist angle, completely flat bands will form. This mechanism is applicable to other twisted bilayers of two-dimensional polar crystals.
\end{abstract}

%\pacs{75.50.Dd, 71.70.Fk, 73.20.Pr, 72.25.Dc}
\maketitle
In condensed matters, an unusual characteristic of Bloch electrons is the existence of flat bands. Being weakly dispersive, a flat band has a vanishingly small band width ($W$) and accordingly, high density of states, inducing strong Coulomb interactions ($U$) between electrons, with $U\gg W$. If the flat band is at the Fermi level, the kinetic energy of the electrons that is confined by $W$ is substantially suppressed to be much smaller than the Coulomb interaction. Thus, the associated system may exhibit pronounced correlation effects~\cite{luttinger} as already seen in various exotic quantum states. These include superconductivity~\cite{miyahara}, ferromagnetism~\cite{congjun}, Wigner crystal~\cite{dassarma}, and  zero-magnetic field fractional quantum Hall effects (QHE) of Bloch states~\cite{frac1,frac2,frac3,frac4,frac5}. Since flat bands provide a route to accessing correlated electronic states, searching for new materials with flat bands is important and currently under active investigations.

Recent theoretical and experimental advances have shown that such flat bands could be obtained in twisted van der Waals (vdW) heterostructures assembled from atomically thin two-dimensional (2D) crystals. Due to the twist-induced misalignment between constituent layers, a twisted vdW heterostructures will develop complex lateral morphologies usually showing as a moir\'{e} pattern with  periodicity much longer than the interatomic distance. This special moir\'{e} superlattice creates strong modulation on the electronic interlayer coupling, leading to interesting physics such as fractional QHE~\cite{moire1,moire2,moire3}, gap opening~\cite{moire4}, and moir\'{e} excitons~\cite{moire5,moire6,moire7,moire8}. Particularly, the electronic structure of twisted graphene bilayers (TBG) can be tailored to develop isolated flat bands at some magic angles~\cite{flat1,flat2,flat3,flat4}. Furthermore, it has also been shown that the value of magic angles relies on the interlayer coupling that can be tuned by varying the interlayer spacing with hydrostatic pressure~\cite{flat5,flat6}.  Experiments have shown that these flat bands are the key to achieve the correlated insulating and superconductive phases in graphene systems~\cite{exp1,exp2,exp3}. Nearly flat bands and possible correlated effects are also predicted for twisted bilayers of transition-metal dichalcogenides (TMDs) at specific twist angles~\cite{tmd1,tmd2}.

In this Letter, we investigate the electronic properties of bilayers and twisted bilayers of hexagonal boron nitride ($h$BN), a typical two-component 2D polar  crystals constituting cation (B) and anion (N) atoms with quantum mechanical simulations.  Our results reveal a new mechanism for the formation of Bloch flat bands other than the known mechanism for twisted graphene bilayer that needs a magic angle.  We show that in the bilayer form of $h$BN, the band edge states are sensitive to the stacking patterns of cation atoms and anion atoms between layers. Namely, the valence band edge will shift upwards if the anion atoms in one layer are on the top of those in the other layer; while the conduction band edge will shift downwards if the cation atoms in one layer are on the top of those in the other layer. Accordingly, in the twisted bilayer form of $h$BN, the valence band maximum (VBM) states are confined to the regions where anion atoms in both layers are on top of each other; while the conduction band minimum (CBM) states are confined to the regions where cation atoms in both layers are on top of each other. Because regions with a specific stacking pattern are spatially separated in the moir\'{e} supperlattice, especially at sufficiently small twist angles, the VBM and CBM states are actually localized, forming the expected flat bands.

Our simulations are carried out with a  density-functional based tight-binding (DFTB) theory with a local orbital basis~\cite{dftb1,dftb2}, which has been widely applied successfully to various forms of $h$BN including sheets~\cite{bn1} and nanotubes~\cite{bn2}. A direct comparison of the electronic structures calculated by DFTB and by first-principles approach ~\cite{vasp} for $h$BN bilayers demonstrates the validity of DFTB in dealing with $h$BN for the main purpose of the present study, see an analysis in the Supplementary Materials.

\begin{figure}[tb]
\includegraphics[width=0.9\columnwidth]{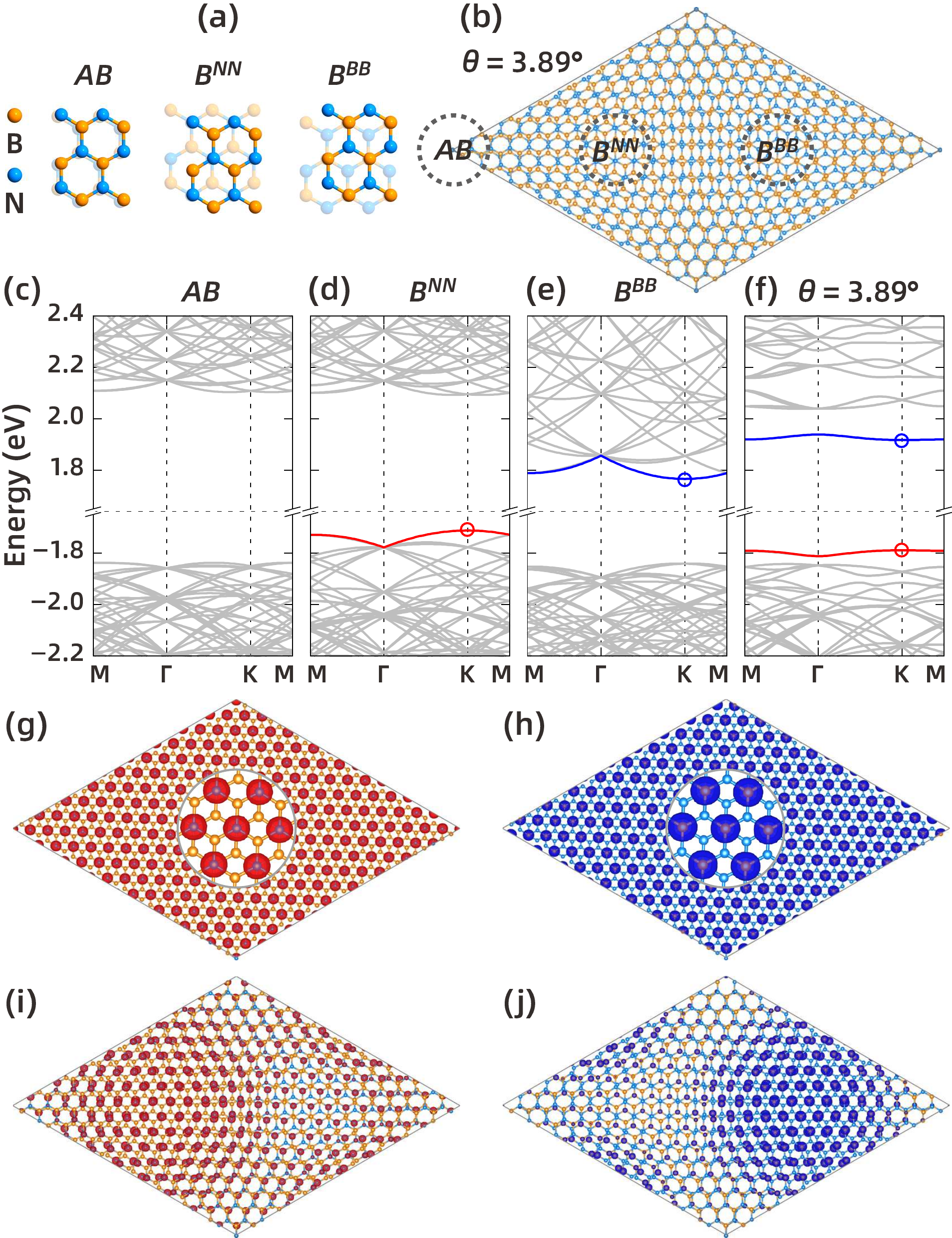}
\caption{(color online)
(a) $h$BN bilayers with $AB$ stacking and two Bernal stackings.  (b) Twisted $h$BN bilayer with a twist angle $\theta=3.89^{\circ}$.
Energy bands of $h$BN bilayers with (c) $AB$ stacking, (d) $B^{NN}$ Bernal stacking, and (e) $B^{BB}$ Bernal stacking.
(f) Energy bands of a twisted $h$BN bilayer with a twist angle $\theta=3.89^{\circ}$.
The dashed horizontal line indicates the zero energy level of the $h$BN bilayer with $AB$ stacking.
(g) Electronic density distribution of the valence band edge state at $k=K$ as marked by a circle in (d).
(h) Electronic density distribution of  the conduction band edge state at $k=K$ as marked by circle in (e).
Inserts in (g) and (h) zoom in the atomic configuration and the distribution of charge on atoms.
Electronic density distribution of (i) the valence band edge state and (j) the  conduction band edge state at $k=K$, as marked by circles in (f).  }%
\label{field}%
\end{figure}

To illustrate the stated idea to find flat bands in 2D, it is appropriate to use twisted hexagonal boron nitride ($h$BN) bilayer as a representative system because $h$BN is a typical 2D polar crystal and its electronic property of monolayer $h$BN has been well investigated experimentally and theoretically. Different from graphene, monolayer $h$BN adopts an energy gap due to the symmetry breaking of $A-B$ sublattice of the hexagon lattice, where the large polarization between cation B and anion N atoms enlarges the energy gap, driving $h$BN into an insulator. The ground state of $h$BN bilayer assumes a $AB$ stacking where B atoms of one layer are on top of N atoms of the other layer and $vice$ $versa$, Fig.~1(a)[left]. If twisted for a small angle $\theta$, the resulting moir\'{e} supperlattice as shown in Fig.~1(b) can also develop two additional Bernal stackings, having N atoms (B atoms) of one layer on the top of N atoms (B atoms) of the other layer, labeled as  $B^{NN}$ ($B^{BB}$) stacking, see Fig.~1(a)[middle] and [right].

Focusing on the electronic structures, we first calculate the energy bands of $h$BN bilayer considering $AB$, $B^{NN}$, and $B^{BB}$ stackings. Results are shown in Figs.~1(c), (d) and (e). Note that for the convenience to compare with twisted bilayer form, these calculations are performed with a cell size comparable with the  moir\'{e} supperlattice at the twist angle $\theta=3.89^{\circ}$, see Fig.~1(b). Interestingly, the electronic states for different stacking patterns display special energy orders. Compared to $AB$ and $B^{BB}$ stackings, the valence band edge of $B^{NN}$  stacking is of higher energy, Fig.~1(d); while compared to $AB$ and $B^{NN}$ stackings, the conduction band edge of $B^{BB}$ stacking is lower, Fig.~1(e).

\begin{figure}[tb]
\includegraphics[width=1\columnwidth]{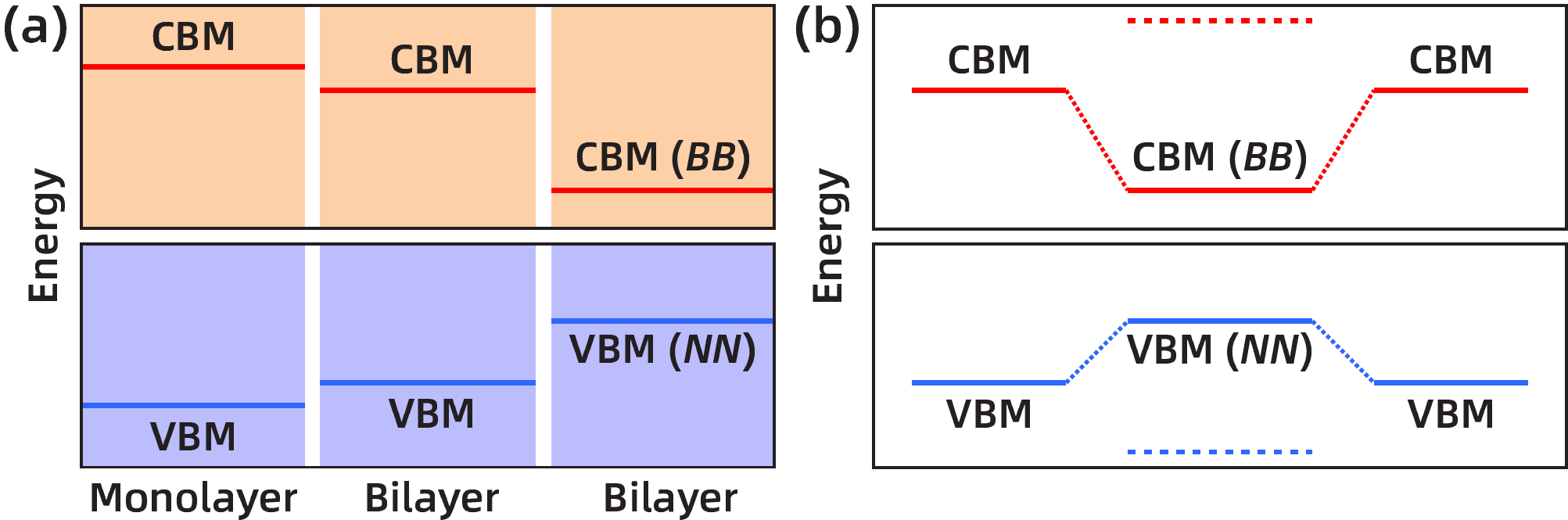}
\caption{(color online)
(a) Energy-level diagrams of VBM and CBM  for monolayer and bilayer $h$BN with different stackings neglecting the energy dispersions. (b) Energy level diagrams showing the splitting of VBM and CBM states due to stacking effects.
}%
\label{bond}%
\end{figure}

To understand these intriguing phenomena caused by the different stacking patterns between layers, we plot schematically the evolution of the band edge states of various layered $h$BN. Fig.~2(a)[left] shows the VBM state and the CBM state of monolayer $h$BN. When forming bilayer, the reduced quantum confinement (without considering the stacking effects) causes slightly a upshift of VBM states and a downshift of CBM states, Fig.~2(b)[middle]. Next, to show the stacking effect, we employ the Hamiltonian
\begin{equation} \label{h12}
H_{12}=\left( \begin{array}{cc}{\varepsilon_{1}} & { U_{12}} \\ {U_{12}^{\dag}} & {\varepsilon_{2}} \end{array}\right)
\end{equation}
where $\varepsilon_{1}$ and $\varepsilon_{2}$ denote the states of layer 1 and layer 2, respectively.  $ U_{12}$  depicts the interlayer coupling.  It is worth noting that the energy levels of valence states of cation B atom is much higher than those of the anion N atom. Thus, for $h$BN, the occupied VBM states are around N atoms and the empty CBM states reside on B atoms and it is the same for the $h$BN bilayer. For example, inserts of Figs.~1(g) and (h) show that for the $B^{NN}$  stacking, the VBM states are situated at N atoms and for the $B^{BB}$ stacking the CBM states are situated at B atoms. We further note that in the $B^{NN}$  stacking, the interlayer coupling $ U_{12}$ essentially relies on the interaction between the N atoms in one layer and the N atoms in other layer, and $\varepsilon_{1}=\varepsilon_{2}$ are VBM states for both layers. This indicates,
\begin{equation} \label{eq2}
\varepsilon_{\pm}=\varepsilon_{1}\pm  U_{12}.
\end{equation}
The result is schematically shown in Fig.~2(b)[lower panel], giving rise to higher energy of the bilayer VBM states, Fig.~2(a)[right]. On the other hand, for the $B^{BB}$ stacking, the interlayer coupling ($U_{12}$) is between the {CBM} states on B atoms in both layers. Again, $\varepsilon_{1}=\varepsilon_{2}$. Hence, similar result of Eq.~(\ref{eq2}) can be obtained, Fig.~2(b)[upper panel], indicating a downshift to lower energy of the bilayer CBM states, Fig.~2(a)[right].  However, for the $AB$ stacking, the stacking effect on CBM and VBM states is much less, where the interlayer coupling ($U_{12}$) is between the CBM states on B atom in one layer ($\varepsilon_{1}$) and the VBM states on N atoms in other layer ($\varepsilon_{2}$). Notice that $U_{12}\ll\varepsilon_{1}-\varepsilon_{2}$, we have,
\begin{equation} \label{eq3}
\varepsilon_{+}=\varepsilon_{1}+ U_{12}^2/(\varepsilon_{1}-\varepsilon_{2});\varepsilon_{-}=\varepsilon_{2}- U_{12}^2/(\varepsilon_{1}-\varepsilon_{2}),
\end{equation}
showing that the variations in the CBM and VBM states are small.

Because the $B^{NN}$  stacking has the highest VBM states and the $B^{BB}$ stacking has the lowest CBM states than other stacking patterns, the VBM and CBM states of the twisted $h$BN bilayer are expected to reside within the $B^{NN}$  stacking  region and the $B^{BB}$ stacking  region, respectively, of the  moir\'{e} supperlattice.  As a demonstration, Figs.~1(i) and (j) displaying the charge density distributions of both VBM and CBM states  at high symmetry $k=K$ point for the considered twisted  $h$BN bilayer, Fig.~1(f), show that indeed the VBM  states are within the $B^{NN}$ stacking region and the CBM states are confined to the $B^{BB}$  stacking  region.

\begin{figure}[tb]
\includegraphics[width=0.9\columnwidth]{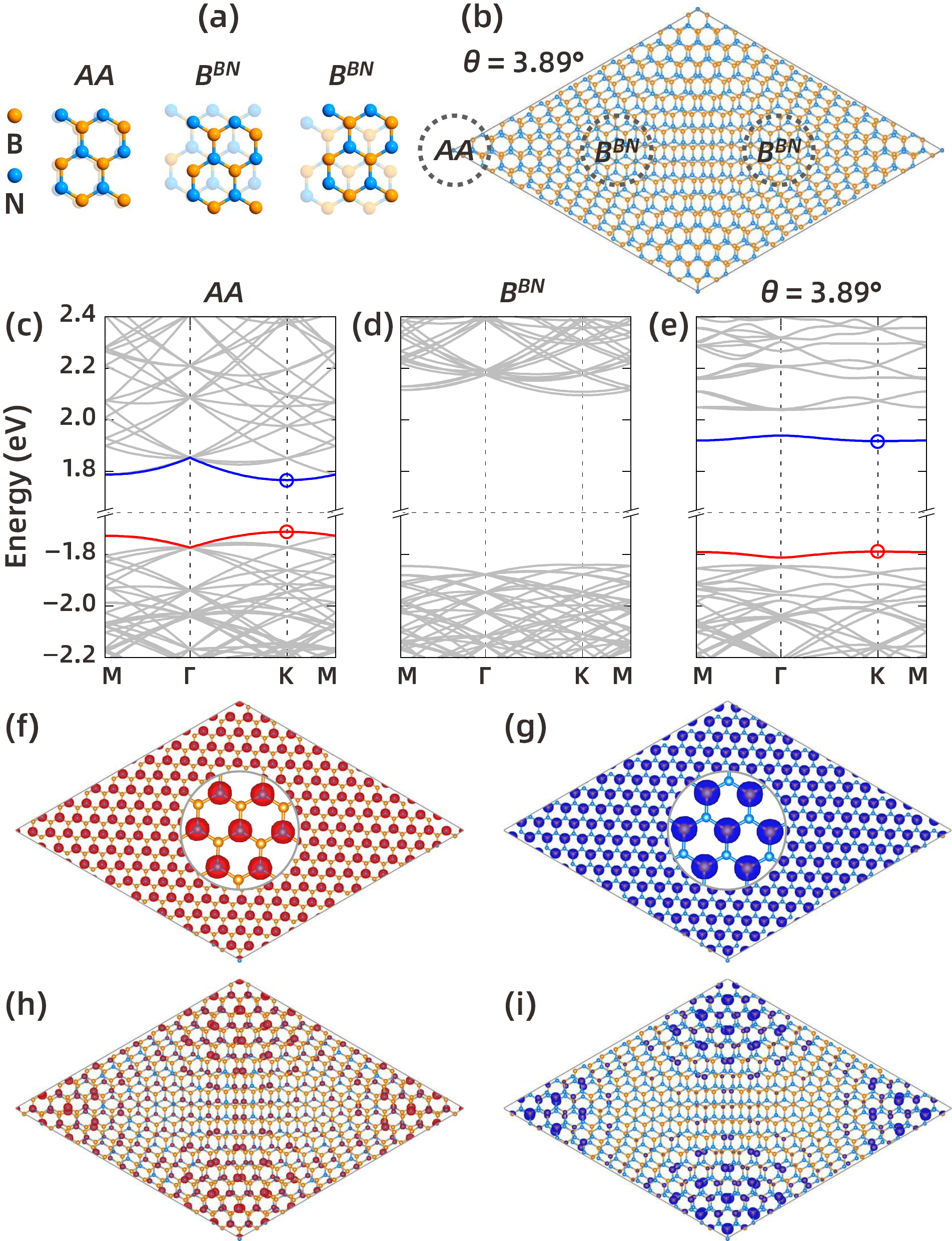}
\caption{(color online)
(a) $h$BN bilayers with $AA$ stacking and a Bernal stacking $B^{BN}$.  (b) Twisted $h$BN bilayer with a twist angle $\theta=3.89^{\circ}$.
Energy bands of $h$BN bilayers with (c) AB stacking and (d) $B^{BN}$ Bernal stacking. (e) Energy bands of a twisted $h$BN bilayer
with a twist angle $\theta=3.89^{\circ}$. The dashed horizontal line indicates the zero energy level of the $h$BN bilayer with $AA$ stacking.
(f) Electronic density distribution of  the valence band edge state at $k=K$ as marked by circle in (c). (g) Electronic density distribution of
the conduction band edge state at $k=K$ as marked by circle in (c).  Inserts in (f) and (g) zoom in the atomic configuration and the distribution of the charge.
Electronic density distribution of (h) the valence band edge state and (i) the conduction band edge state at $k=K$ as marked by circles in (e).  }%
\label{field}%
\end{figure}

For a polar 2D crystal, its twisted bilayer form can be also obtained by applying a small twist angle to the bilayer form with $AA$ stacking, where the B atoms of one layer are on top of the B atoms of the other layer and the same for N atoms, see Fig.~3(a)[left]. The resulting moir\'{e} supperlattice as shown in Fig.~3(b) develops a new Bernal stacking besides the $AA$ stacking, having N atoms (B atoms) of one layer are on the top of B atoms (N atoms) of the other layer, labeled as  $B^{BN}$, see Fig.~3(a)[right].

Figs.~3(c) and (d) displaying the energy bands of $h$BN bilayer for these two stacking patterns show that compared to the $B^{BN}$ stacking, the  valence (conduction) band edge of $AA$ stacking is higher (lower). These results can also be understood by using the Hamiltonian of Eq.~(\ref{eq2}).  For the $AA$ stacking, the interlayer interaction is not only in between the B atoms of both layers but also in between the N atoms of both layers. Therefore, for the VBM states that reside on N atoms in both monolayers, as shown in Fig.~3(f), the coupling between the N states shifts the bilayer VBM states up. Similarly, CBM states  reside on B atoms, as shown in Fig.~3(g), the coupling between B states push down the bilayer CBM states. Note that the stacking of  the $B^{BN}$ pushes the CBM up and and VBM down in energy, as inferred by Eq.~(\ref{eq3}), so it has little contribution to the charge distribution at the CBM and VBM edges of the twisted $h$BN bilayer. That is, both the CBM  and VBM states of the twisted $h$BN bilayer shown in Fig.~3(e) reside within the $AA$ stacking region, as demonstrated by Figs.~3(h) and (i).

\begin{figure}[tb]
\includegraphics[width=1\columnwidth]{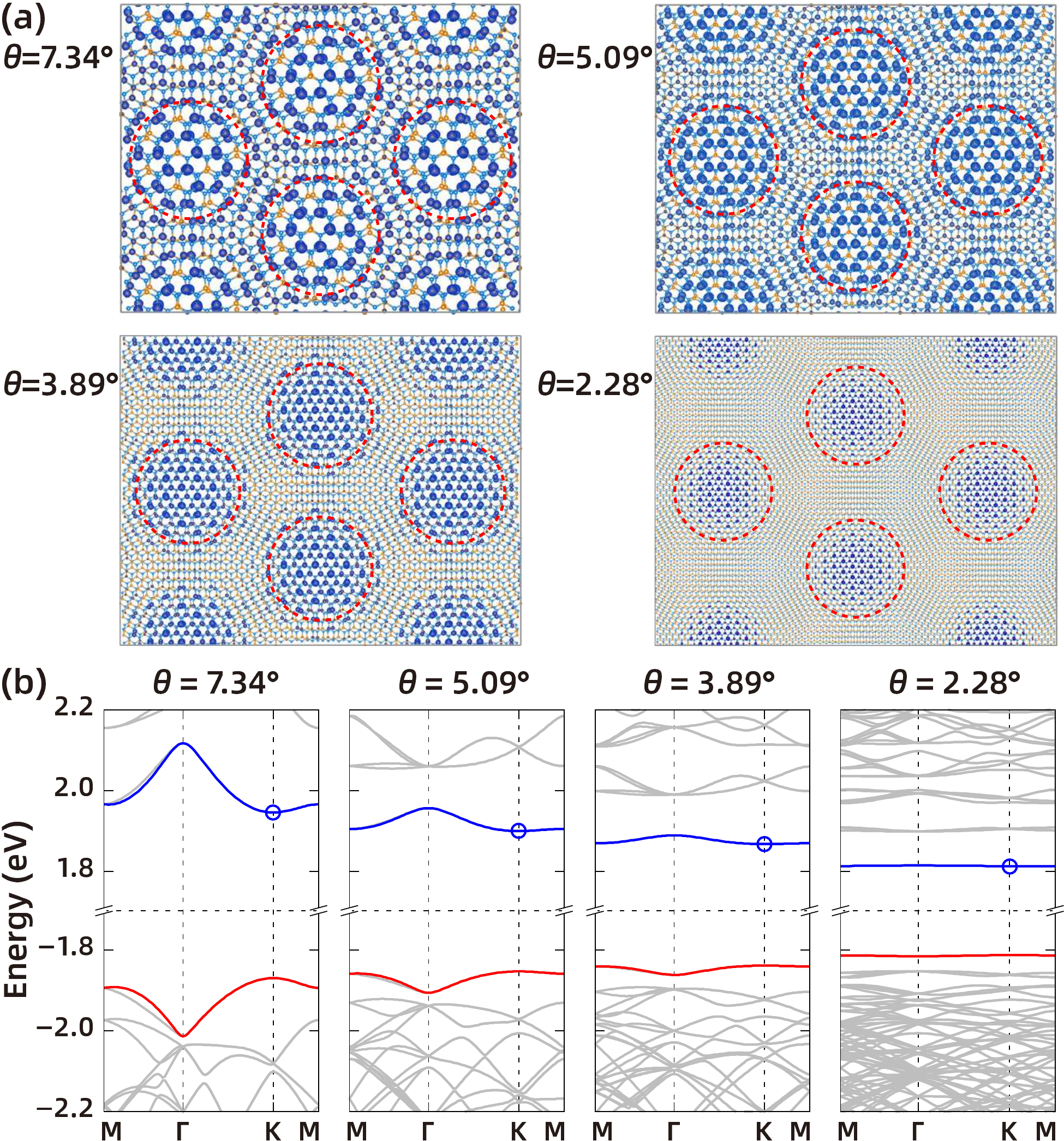}
\caption{(color online)
(a) Twisted $h$BN bilayers with different twist angle $\theta$. The circles with radius $R$ highlight the $AA$ stacking regions.
Electronic density distribution of the  conduction band edge state at $k=K$, as marked by circles in (b). (b) Energy bands of twisted $h$BN bilayers at different twist angle $\theta$. The dashed horizontal line indicates the energy zero. }%
\label{field}%
\end{figure}

Because for the moir\'{e} supperlattice, regions for a specific stacking pattern are spatially separated, it is thus possible that the electronic states (CBM states and/or VBM states) resided in these regions form flat energy bands when the separation of the regions is large. Indeed, as shown in Fig.~4(a), the charge density distribution of the CBM state at $k=K$ of the twisted $h$BN bilayers  displays a gradual localization around the $AA$ stacking regions with the decrease of the twist angle $\theta$. Similar trend can be also identified for VBM states as shown in Fig.~S3 of the Supplemental Material. This is mainly because that the separation between different $AA$ stacking regions is larger at smaller $\theta$ such that the overlap between states in these regions is less. It is important to note that the stacking patterns in a twisted bilayer is not as perfect as that in the bilayer form. This deviation can be characterized with an in-plane displacement, $d$, between the two atoms that are on top of each other in the bilayer form. Hence, it is proper to introduce a critical displacement $d_c$ where a stacking pattern can be reasonably identified for a region satisfying  $d<d_c$ and the radius of each region is simply $R=Za_{BN}$, where $a_{BN}$ is the lattice constant of the monolayer~\cite{latt1} and integer $Z = \text {INT}(d_{c}/a_{BN}\theta)+1$. With $d_c=a_{BN}/2\sqrt{3}$, i.e., half of the B-N bond length, Fig.~4(a) shows the $AA$ stacking regions highlighted by circles for different twist angles $\theta$. Notice that the distance between the centers of two regions are actually the lattice constant of the moir\'{e} supperlattice, $L=a_{BN}/[2(1-\cos\theta)]^{1/2}$~\cite{latt2}. In this way, the space between regions for each stacking pattern can be approximately described with a simple relation of $L-2R\simeq a_{BN}(1-1/\sqrt{3})/\theta$. Apparently, the space between regions can be huge at small $\theta$. For example, $L-2R\simeq 2.6$~nm at $\theta=2.28^{\circ}$, Fig.~4(a). Such a separation hints that the coupling between electronic states resided in these regions is weak, giving rise to flat bands. Indeed, Fig.~4(b) shows that for twisted $h$BN bilayers, the band width of both VBM and CBM states become narrower when $\theta$ is reduced. At $\theta=2.28^{\circ}$, the VBM and CBM bands are already flat with band widths only about $3$~meV. We  note that structural relaxation has little effect on the formation of the obtained flat bands as illustrated by Fig.~S2 of the Supplemental Material.

The obtained flat bands in twisted $h$BN bilayers are important and the above results reveal several important aspects which should be common for  polar 2D crystals.
(i) In the bilayer form, the  valence band edge and conduction band edge are sensitive to the interlayer stacking of the anions (e.g, N) and the cations (e.g, B), respectively. (ii) In the twisted bilayer structure, the band alignment between different stacking regions follows the order of the valence or conduction band edge states of the different stacking patterns forming the bilayer.  (iii) Most importantly, the emerged flat bands are found to be an intrinsic property for twisted bilayers of polar 2D crystal with large band gaps. Given a sufficiently small twist angle $\theta$, regions of a particular stacking pattern for a moir\'{e} supperlattice can be well separated. As a result, electronic states preferentially  resided in these regions are spatially localized and form flat bands. These aspects indicate that the mechanism towards Bloch flat bands revealed here is different from the one of twisted graphene bilayers that needs a magic angle.

In summary, using the DFTB method and taking $h$BN as an example, we have studied the electronic proprieties of the bilayer and twisted bilayers of polar crystals and shown that in the twisted bilayer structure, Bloch flat bands emerge as long as the twist angle is sufficiently small. By analyzing the evolution of the valence and conduction band edges of the bilayer with different stacking patterns, a mechanism attributed to the splitting of the band edge states induced by different stacking patterns is revealed. This mechanism differs from that for twisted graphene bilayers where a magic angle is needed. This new mechanism brings up two benefits. (i) Similar flat bands are expected to emerge in the twisted bilayer of other polar 2D crystals such as transition-metal dichalcogenides~\cite{tmd2,polar1}, and 2D oxides~\cite{polar2} at small twist angle $\theta$. (ii) The possibility of the formation of the flat bands for a twisted bilayer can be inferred by just examining the energy order of the VBM and CBM states of the various bilayers forming the stacking patterns of the twisted bilayer. The direct simulation of the twisted bilayer, which is computationally demanding, could be avoided. These findings, thus pave a new route to explore Bloch flat bands and associated many-body physics in 2D materials.

This work was supported by NSFC under Grants Nos. 11674022, 51672023, 11634003, and U1530401. D.-B.Z. was supported by the Fundamental Research Funds for the Central Universities. X.-J.Z. was supported by Postdoctoral innovative talents support program (BX201700025).

\end{document}